\documentclass[aps,twocolumn,prb]{revtex4-2}

\usepackage{graphicx}
\usepackage{dcolumn}
\usepackage{bm}

\begin{document}

\title{Quasi-resonant diffusion of wave packets in one-dimensional disordered mosaic lattices}

\author{Ba Phi Nguyen}
\email{nguyenbaphi@muce.edu.vn}
\affiliation{Department of Basic Sciences, Mientrung University of Civil Engineering, Tuy Hoa 620000, Vietnam}
\affiliation{Research Institute for Basic Sciences, Ajou University, Suwon 16499, Korea}
\author{Duy Khuong Phung}
\affiliation{Computing Fundamentals Department, FPT University, Hanoi 100000, Vietnam}
\author{Kihong Kim}
\email{khkim@ajou.ac.kr}
\affiliation{Department of Physics, Ajou University, Suwon 16499, Korea}
\affiliation{School of Physics, Korea Institute for Advanced Study, Seoul 02455, Korea}

\begin{abstract}
We investigate numerically the time evolution of wave packets incident on one-dimensional semi-infinite lattices with mosaic modulated random on-site potentials, which are characterized by the integer-valued modulation period $\kappa$ and the disorder strength $W$.
For Gaussian wave packets with the central energy $E_0$ and a small spectral width,
we perform extensive numerical calculations
of the disorder-averaged time-dependent reflectance, $\langle R(t)\rangle$,
for various values of $E_0$, $\kappa$, and $W$.
We find that the long-time behavior of
$\langle R(t)\rangle$ obeys a power law of the form $t^{-\gamma}$ in all cases.
In the presence of the mosaic modulation, $\gamma$ is equal to 2 for almost all values of $E_0$, implying the onset of the Anderson
localization, while at a finite number of discrete values of $E_0$ dependent on $\kappa$, $\gamma$ approaches 3/2, implying the onset of the classical diffusion.
This phenomenon is independent of the disorder strength and arises in a quasi-resonant manner such that $\gamma$ varies rapidly from 3/2 to 2
in a narrow energy range as $E_0$ varies away from the quasi-resonance values.
We deduce a simple analytical formula for the quasi-resonance energies and provide an explanation of the delocalization phenomenon based on the interplay between randomness and band structure
and the node structure of the wave functions.
We explore the nature of the states at the quasi-resonance energies using a finite-size scaling analysis
of the average participation ratio and find that the states
are neither extended nor exponentially localized, but ciritical states. 
\end{abstract}

\maketitle

\section{\label{sec:level1} Introduction}

Anderson localization of classical waves and quantum particles occurs due to the interference of multiply scattered waves in spatially random media. Since it was discovered theoretically over 60 years ago by Anderson, it has been studied extensively in many areas of physics \cite{And,Lee,Eve}. Anderson localization arises universally for all kinds of waves and
many aspects of the phenomenon have been explored and understood in detail \cite{Hu,Mod,Gre,Seg}.
Nevertheless, there still exist features which are not fully understood and
new aspects of localization continue to be discovered
when new elements are included in the system \cite{Rez,Ley,Der,Seg2,Kim7,Pel}.

One of the prominent results of early theories of localization is that in the simplest
one-dimensional (1D) and two-dimensional random systems, all eigenstates are exponentially localized
even in the presence of infinitesimally weak disorder \cite{Abr}. However, it has been found that this conclusion is not always true
in more general random systems and there exist various situations where some states are not exponentially localized, but
are either extended or critically localized. Representative examples include the cases where distinct
kinds of impedance matching phenomena such as the Brewster anomaly \cite{Sip,Kim1,Kim2,Jor,Kim6,Kim3}
and the Klein effect \cite{Zha,Cha,Kim4,Kim5} happen or the random potential is spatially correlated
with short- or long-range correlations \cite{Dun,Bell,Wu,Izr,Maj0,Maj,Mou1,Izr1,Kan,Kuh}.
For instance, it has been demonstrated both theoretically and experimentally that there exist
extended states at two energy values in the 1D
random dimer model which contains a special type of short-range correlated disorder \cite{Dun,Bell}.
This model has been generalized to the case of random $N$-mer systems and analytical expressions for the values
of the resonant energies at which delocalization arises have been obtained \cite{Wu,Izr,Maj0,Maj}.

A discrete set of delocalized states can appear in short-range correlated random systems.
In contrast, a continuum of delocalized states with sharp mobility edges was predicted theoretically and confirmed experimentally to appear in long-range correlated random systems in 1D \cite{Mou1,Izr1,Kan,Kuh,San1,San2,San3}.
The mobility edges, which are the energy values separating localized and extended states, can also be present
in a wide range of quasiperiodic models in 1D \cite{Bid2,Pur,Lus,An,San4,Wan,Zen,Liu,Zen1,Gon,Wang,Dwi}.
Recently, an exactly solvable 1D model with multiple mobility edges called quasiperiodic mosaic lattice model has been proposed \cite{Wan}.
In that model, a quasiperiodic on-site potential exists only at periodically spaced sites, while the potential is constant
at all other sites. The number and the positions of the mobility edges have been shown to depend sensitively on the modulation period $\kappa$.

The main aim of the present paper is to propose a new way of inducing delocalized states in 1D random systems. We
consider a random version of the quasiperiodic mosaic lattice model, which we call disordered mosaic lattice model,
where the on-site potential takes a random value only at equally spaced sites with a period $\kappa$.
We study the transport and localization properties of the proposed model primarily by
investigating the time evolution of wave packets incident on an effectively semi-infinite disordered mosaic lattice chain.
More specifically, we calculate the time-dependent reflectance averaged over
a large number of independent disorder configurations, $\langle R(t)\rangle$, for various values of the modulation period $\kappa$ and the central energy of the wave packet, $E_0$. This approach based on the reflection geometry has  substantial experimental advantages over those based on the transmission geometry.

We are especially interested in exploring the long-time scaling behavior of  $\langle R(t)\rangle$, which
obeys a power-law decay of the form $t^{-\gamma}$ for all values of the parameters.
From many previous researches, it has been solidly established that in the cases
where the standard Anderson localization occurs, the exponent $\gamma$
is equal to 2, while in those where the classical diffusion occurs, it is 3/2 \cite{Whi,Schu,Tit,Joh,Ski1,Ski2,Dou,Abu,Ski}.
From extensive numerical calculations, we will find that
in the presence of the mosaic modulation,
$\gamma$ is equal to 2 for almost all values of $E_0$, while at a finite number of discrete values
of $E_0$ dependent on $\kappa$, $\gamma$ approaches 3/2.
In other words, although most eigenstates of the disordered mosaic lattice model are exponentially localized, there appear a finite number of discrete states that are not exponentially localized but display transport behavior characteristic of classical diffusion. We will also find that this phenomenon is independent of the disorder strength and occurs in a quasi-resonant manner.
From the numerical results, we will deduce a simple analytical formula for
the quasi-resonance energies at which delocalization occurs and provide an explanation of the phenomenon
based on the interplay between randomness and band structure and the node structure of the wave functions.
We also explore the nature of the states at the quasi-resonance energies using a finite-size scaling analysis
of the average participation ratio and find that the states
are neither extended nor exponentially localized, but ciritical states.

The rest of this paper is organized as follows. In Sec.~II, we introduce the 1D disordered mosaic lattice model characterized by the time-independent Schr\"odinger equation within the nearest-neighbor tight-binding approximation. We also describe the numerical calculation method and the physical quantities of interest. In Sec.~III, we present the numerical results and discuss the mechanism for the onset of the diffusive behavior. We also explore the nature of the states at the quasi-resonance energies using a finite-size scaling analysis
of the average participation ratio. Finally, in Sec.~IV, we conclude the paper.

\section{Theoretical model and method}

\subsection{Model}

To describe 1D non-interacting spinless particle systems,
we use the standard single-chain tight-binding model
\begin{eqnarray}
J(\psi_{n-1}+\psi_{n+1})+\varepsilon_{n}\psi_{n}= E{\psi_{n}},
\label{equation1}
\end{eqnarray}
where $\psi_{n}$ and $\varepsilon_{n}$ are the wave function amplitude and the on-site potential at the $n$-th lattice site respectively.
$E$ is the energy and $J$ is the coupling strength between nearest-neighbor sites.
From now on, we will measure all energy scales in units of $J$ and set it equal to 1.
In this study, we will investigate the transport and localization properties of a model which we call disordered mosaic lattice model.
This model is defined by
\begin{eqnarray}
\varepsilon_{n}=\left\{\begin{array}{l l}
\beta_{n} \in [-W,W], & \quad \mbox{$n=m\kappa$}\\
V_{0}, & \quad \mbox{otherwise}
\end{array}\right.,
\label{equation2}
\end{eqnarray}
where the inlay parameter $\kappa$ representing the period of the mosaic modulation
is a fixed positive integer larger than 1 and $m$ is an integer running from 1 to $N$. Then the total number of sites $L$ is
equal to $\kappa N$. The on-site potential $\beta_n$ at the $m\kappa$-th site is a random variable uniformly distributed in the interval
$[-W,W]$, where $W$ is the strength of disorder. In all other sites, the on-site potential takes a constant value of $V_{0}$.
In \cite{Wan}, the authors have studied a quasiperiodic mosaic lattice model, where $\beta_n$ is a quasiperiodic potential of Aubry-Andr\'e-type \cite{Aub}.
Our model is different from that model in that $\beta_n$ is random instead of quasiperiodic.
We note that if $\beta_n$ is a constant potential different from $V_0$, the model becomes perfectly periodic with period $\kappa$.

\subsection{\label{sec:level2} Method}

Following the procedure given in \cite{Ski}, we first assume that a monochromatic wave of energy $E$ is incident from the left side of the disordered
region and define the amplitudes of the incident, reflected, and transmitted waves $A$, $B$, and $C$ by
\begin{eqnarray}
\psi_{n}=\left\{\begin{array}{l l}
Ae^{iqn}+Be^{-iqn}, & \quad \mbox{$n= 1$},2\\
Ce^{iqn}, & \quad \mbox{$n= L-1$},L
\end{array}\right.,
\label{equation3}
\end{eqnarray}
where $q$ is related to $E$ by the free-space dispersion relation $E=2\cos q$. In the absence of dissipation, the law of energy conservation
$\vert B\vert^2+\vert C\vert^2=\vert A\vert^2$ should be satisfied.
In order to solve Eq.~(\ref{equation1}) numerically, we first fix $\psi_{L-1}$ to 1, then we obtain $C=\exp[-iq(L-1)]$ and $\psi_L=\exp(iq)$.
Knowing the values of $\psi_{L}$ and $\psi_{L-1}$, we can solve Eq.~(\ref{equation1}) iteratively to obtain $\psi_{L-2}$, $\psi_{L-3}$, $\cdots$, $\psi_2$, $\psi_{1}$. Using the definition of $A$ and $B$ given in Eq.~(\ref{equation3}), we can express them in terms of $\psi_1$ and $\psi_2$:
\begin{eqnarray}
A=e^{-2iq}\frac{\psi_{2}-\psi_{1}e^{-iq}}{1-e^{-2iq}},~B=\frac{\psi_{1}e^{iq}-\psi_{2}}{1-e^{-2iq}}.
\label{equation4}
\end{eqnarray}
We are interested in the case where the length of the disordered region is sufficiently large. Then the large portion of the incident wave power
will be reflected. Therefore we focus on the behavior of the reflection coefficient $\tilde{r}(E)$ and the reflectance $\tilde{R}(E)$ defined by
\begin{eqnarray}
&&\tilde{r}(E)=\frac{B}{A}=e^{2iq} \frac{\psi_{1}e^{iq}-\psi_{2}}{\psi_{2}-\psi_{1}e^{-iq}},\nonumber\\
&&\tilde{R}(E)=\left\vert\frac{B}{A}\right\vert^2=\left\vert\frac{\psi_{2}-\psi_{1}e^{iq}}{\psi_{2}-\psi_{1}e^{-iq}}\right\vert^2.
\label{equation5}
\end{eqnarray}

Next we consider a Gaussian wave packet characterized by the spectrum
\begin{eqnarray}
f(E)=\frac{(2\pi)^{1/4}}{\sqrt\sigma}\exp\left[\ -\frac{(E-E_{0})^2}{4\sigma^2} \right],
\label{equation9}
\end{eqnarray}
where $E_0$ is the central energy of the wave packet and $\sigma$ is its spectral width.
$f(E)$ satisfies
\begin{eqnarray}
\frac{1}{2\pi}\int_{-\infty}^{\infty}dE~\left[ f(E)\right]^2=1.
\end{eqnarray}
The time-dependent reflection coefficient $r(t)$ of the incident Gaussian wave packet can be calculated
using
\begin{eqnarray}
r(t)=\frac{1}{2\pi}\int_{-2}^{2}dE~\tilde{r}(E)f(E)\exp(iEt),
\label{equation7}
\end{eqnarray}
where the range of integration is limited to $-2\le E\le 2$, since
the reflection coefficient $\tilde r(E)$ can be defined only inside the band satisfying $E=2\cos q$.
The disorder-averaged time-dependent reflectance $\left\langle R(t) \right\rangle$ is obtained from
\begin{widetext}
\begin{eqnarray}
\left\langle R(t) \right\rangle=\langle \left\vert r(t)\right\vert^2 \rangle=\frac{1}{(2\pi)^2}\int_{-2}^{2}dE_{1}\int_{-2}^{2}dE_{2}~\langle \tilde{r}(E_{1}) \tilde{r}^{\ast}(E_{2}) \rangle f(E_{1})f^{\ast}(E_{2})\exp[i(E_{1}-E_{2})t],
\label{equation8}
\end{eqnarray}
\end{widetext}
where $\langle\cdots\rangle$ denotes averaging over a large number of different disorder configurations.

According to the previous theories of Anderson localization in the time domain \cite{Whi,Tit,Joh,Ski1,Ski2,Ski}, the exponent characterizing the power-law decay of
$\left\langle R(t) \right\rangle$ in the long-time limit provides an important information on the state of the system.
Specifically, it has been established that $\left\langle R(t) \right\rangle$ decays as $t^{-2}$ in the localized regime, while it decays as $t^{-3/2}$ in the regime
of classical diffusion.
Finally, we point out that studying Anderson localization based on the measurements in the reflection geometry
has substantial experimental advantages over more conventional approaches in the transmission geometry, in that the reflectance is often
more easily measurable than the transmittance and there exist situations where measurements in the transmission mode are not possible.
These advantages are especially relevant to the fields such as optics, acoustics, and seismology \cite{Dur,Boa,Sha,Fou}.

\section{\label{sec:level1} Numerical results and discussion}
\subsection{\label{sec:level2} Numerical results}

\begin{figure}
\centering
\includegraphics[width=10cm]{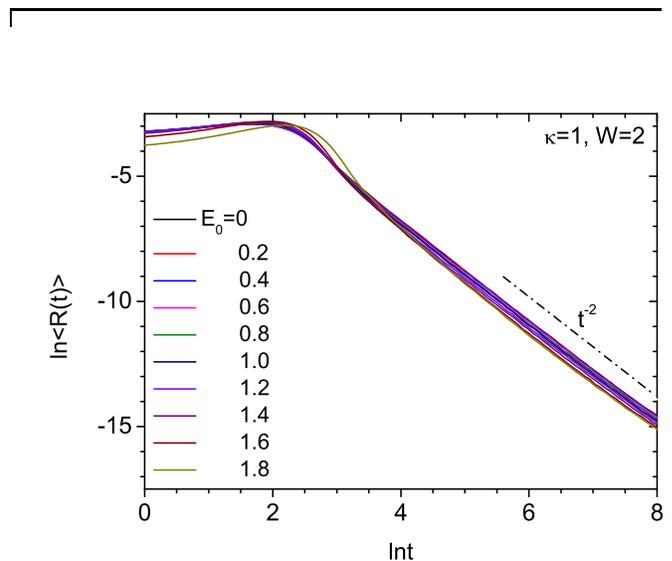}
\caption{Ln--ln plot of the disorder-averaged time-dependent reflectance $\langle R(t) \rangle$ versus time $t$ for various values of $E_{0}$ when $\kappa=1$, $W=2$, $\sigma=0.05$, $V_0=0$, and $L=1000$. The time dependence of $\langle R(t) \rangle$ obeys an inverse-square law of the form $\langle R(t) \rangle \propto t^{-2}$ in the long-time limit.}
\label{fig1}
\end{figure}

\begin{figure}
\centering
\includegraphics[width=10cm]{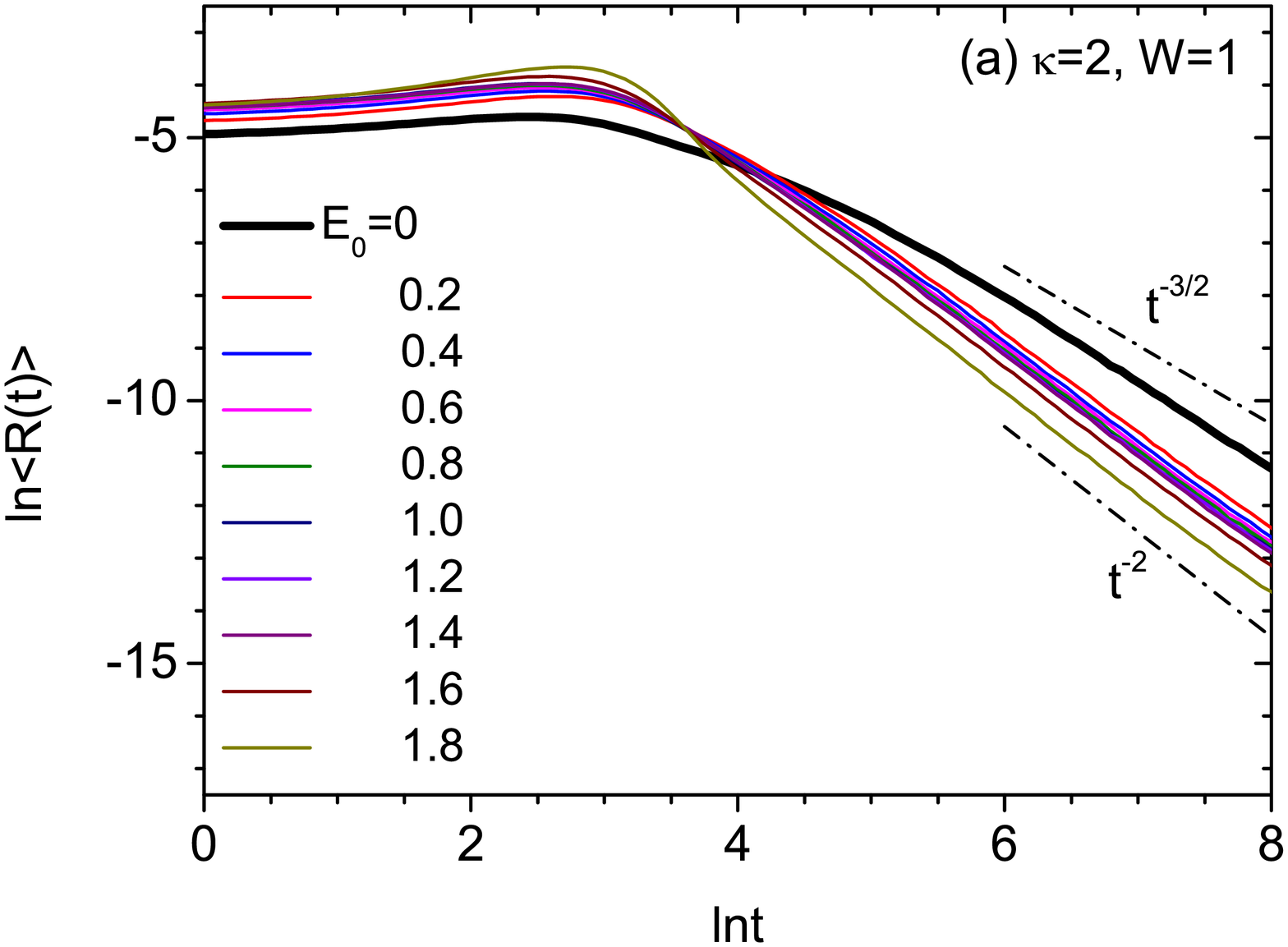}
\includegraphics[width=10cm]{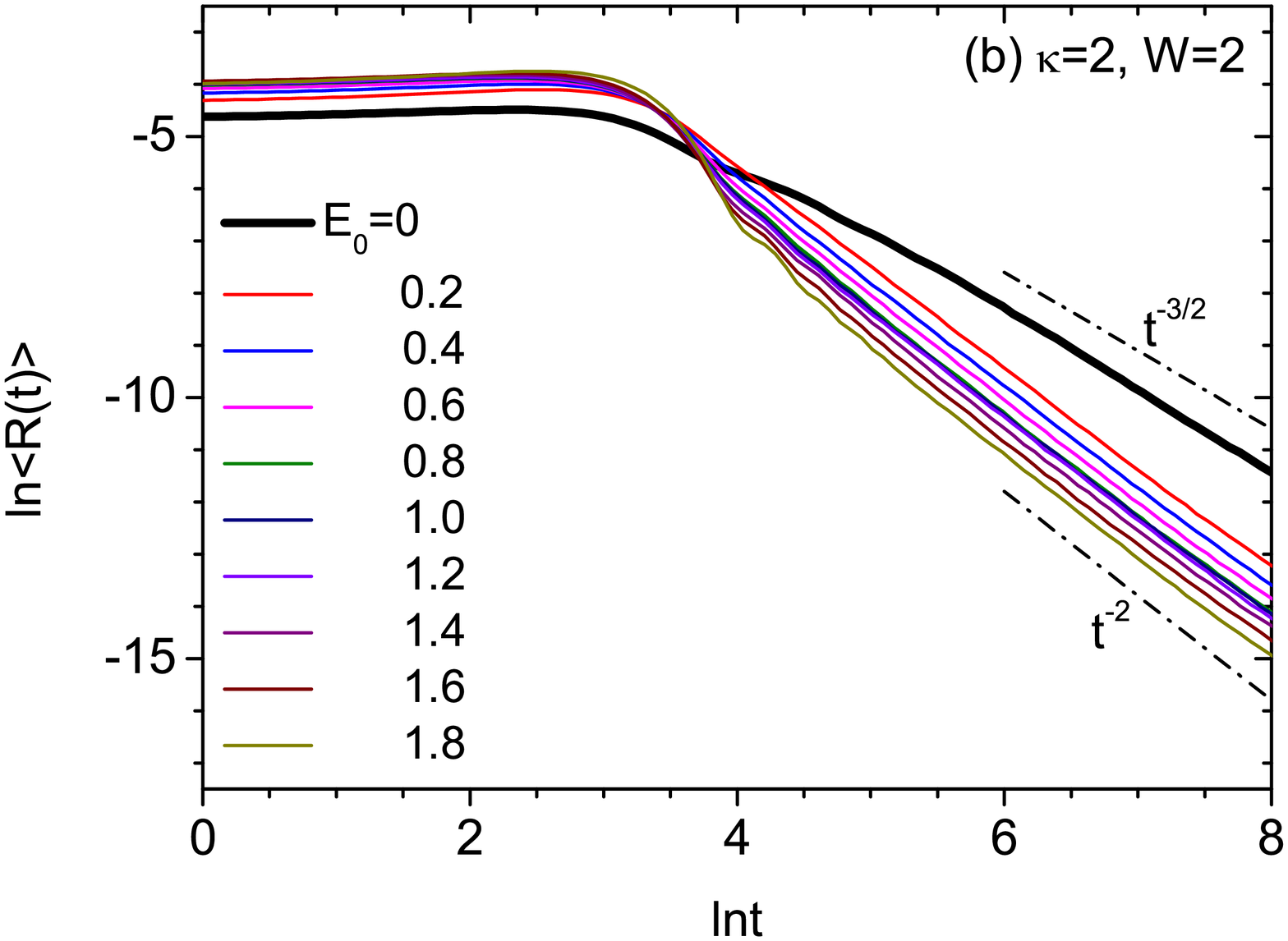}
\includegraphics[width=10cm]{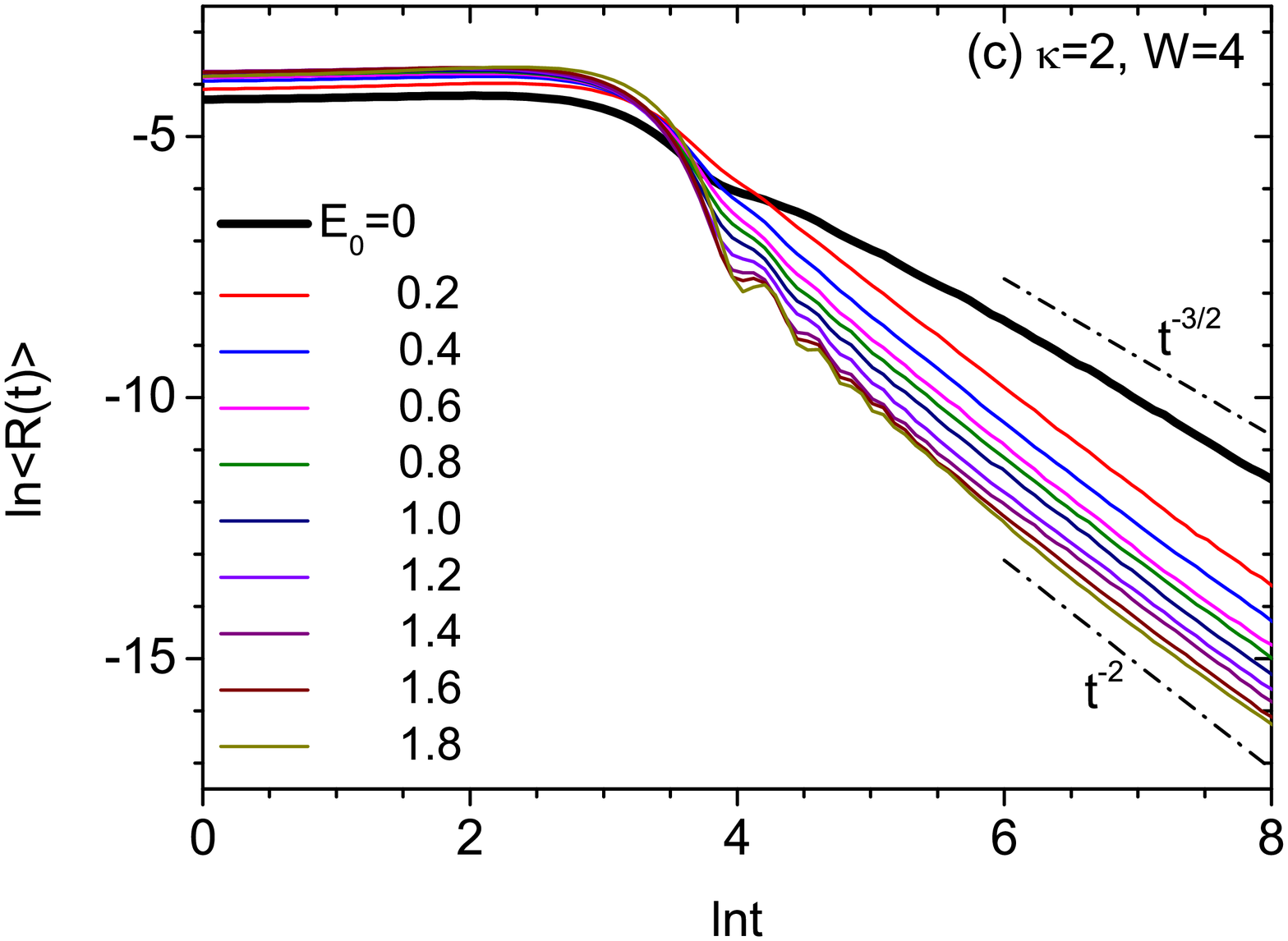}
\caption{Ln--ln plots of $\langle R(t) \rangle$ versus $t$ for various values of $E_{0}$ when $\kappa=2$ and (a) $W=1$, (b) $W=2$, and (c) $W=4$.
$\sigma$ and $V_0$ are fixed to 0.05 and 0 respectively and $L$ is chosen to be 1300 in (a) and 1000 in (b) and (c). The curve for $E_0=0$ that is highlighted with a thick line scales as $t^{-3/2}$ in the long-time limit in contrast with other curves that scale as $t^{-2}$.}
\label{fig2}
\end{figure}

In most of our calculations, we have set $V_{0}=0$ and computed $\langle R(t) \rangle$ by averaging over 10,000 distinct random configurations of $\varepsilon_{n}$. The step size in energy, $\Delta E$, is $10^{-3}$. We have chosen the system size $L$ to be sufficiently large so that
the transmission is negligible and the system can be considered an effectively semi-infinite medium.

Our main goal is to study the dynamics of wave packet propagation in 1D disordered mosaic lattices with $\kappa\ge 2$.
However, it is instructive to first consider the $\kappa=1$ case corresponding to the ordinary Anderson model to clarify the effects of the
mosaic modulation. In a recent study, a numerical analysis of the time-dependent reflectance for the 1D Anderson model
has been presented \cite{Ski}.
It has been reported that for any $W\in[1,4]$ and $E_{0 }\in[-2+2\sigma, 2-2\sigma]$,
the time dependence of $\langle R(t) \rangle$ obeys an inverse-square law of the form $\langle R(t) \rangle \propto t^{-2}$ in the long-time limit.
This implies that all the incident wave packets exhibit an Anderson localization behavior. It has also been shown that this behavior is
independent of the spectral shape of the wave packet or its spectral width $\sigma$ as long as it is as small
as 0.05 and 1. In the present work, we consider Gaussian wave packets with a fixed spectral width of $\sigma=0.05$.
In Fig.~\ref{fig1}, we show our calculations for the time decay of $\langle R(t) \rangle$ when $\kappa=1$, $W=2$, and the system size $L=1000$,
which agree well with \cite{Ski}.

We are interested in exploring the transport behavior
when the random mosaic modulation defined by Eq.~(\ref{equation2}) is introduced into the on-site potential of a lattice model.
In Fig.~\ref{fig2}, we set $\kappa=2$ and plot the time evolution of $\langle R(t) \rangle$ for various values of $E_{0}$
when $W=1$, 2, and 4.
We find that for all values of $E_0$, $\langle R(t) \rangle$ shows a power-law decay of the form $\langle R(t) \rangle \propto t^{-\gamma}$ in the long-time limit. The exponent $\gamma$ is equal to 2 for almost all $E_{0}$ values except for a narrow region close to $E_0=0$.
In contrast, the value of $\gamma$ at $E_0=0$ is equal to $3/2$ with a good approximation and is markedly different from the Anderson localization behavior shown
for other $E_0$ values.
In addition, the value of $\langle R(t) \rangle$ for $E_0=0$ is noticeably larger than those for other $E_0$ values when $\ln t$ is sufficiently large, as can be seen from the curves drawn with thick lines in Fig.~\ref{fig2}.
This behavior occurs for all values of the disorder parameter $W$ considered here, as long as the system length $L$ is
sufficiently large such that the effect due to the leakage of the wave packet into the transmitted region is negligible. When the disorder
is weak, we need to use a larger $L$ to satisfy such a condition.

\begin{figure}
\includegraphics[width=10cm]{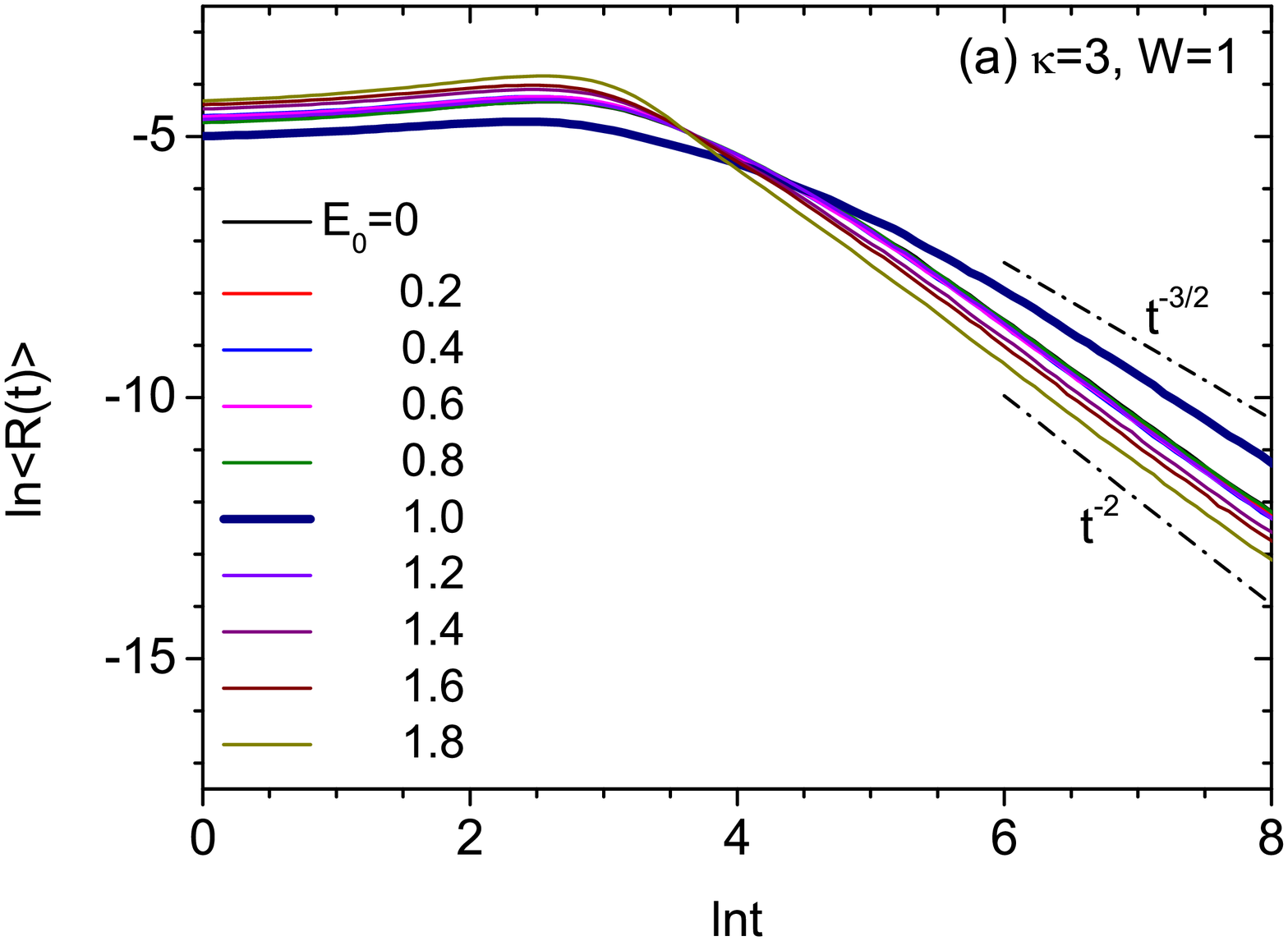}
\includegraphics[width=10cm]{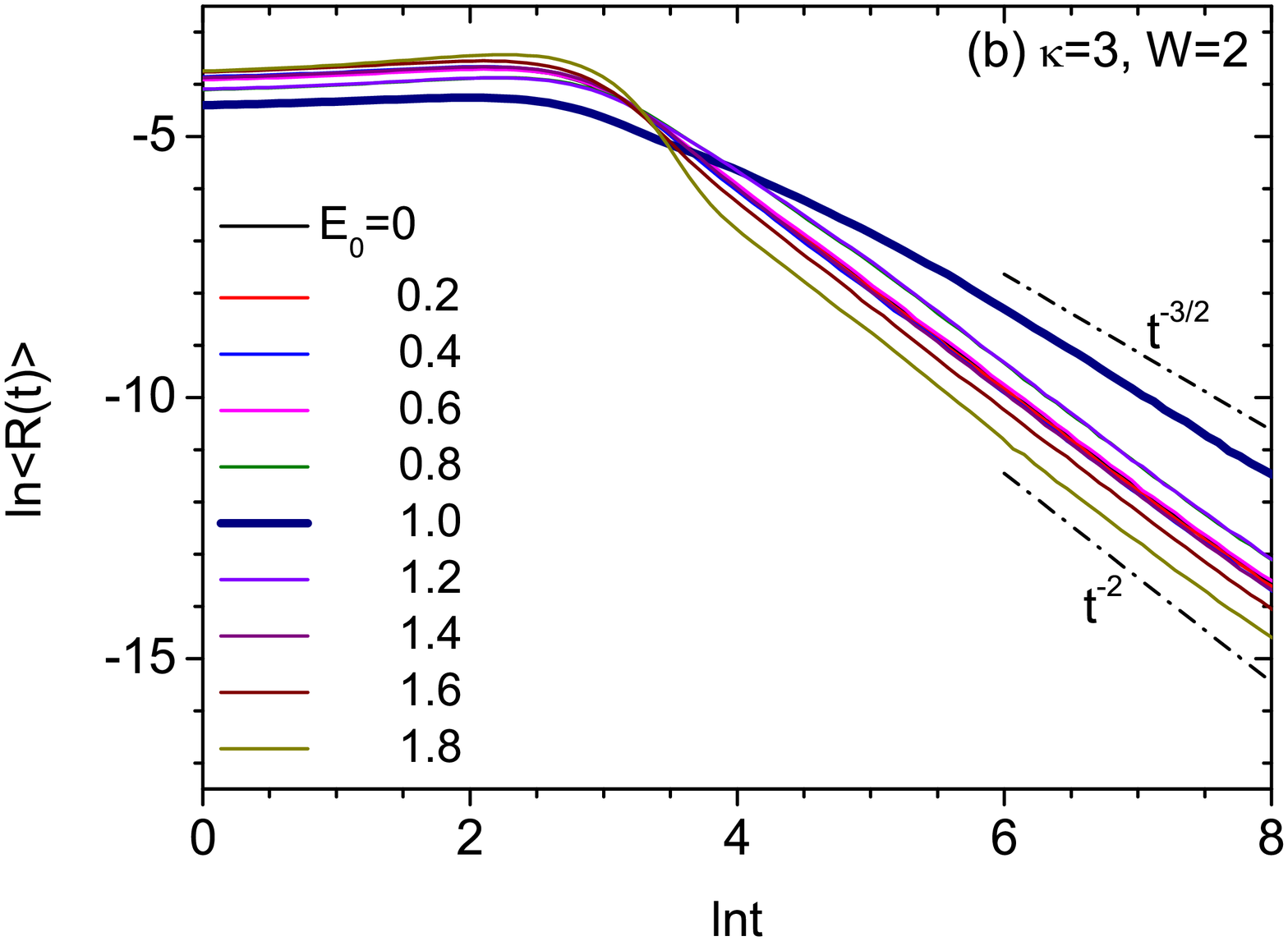}
\includegraphics[width=10cm]{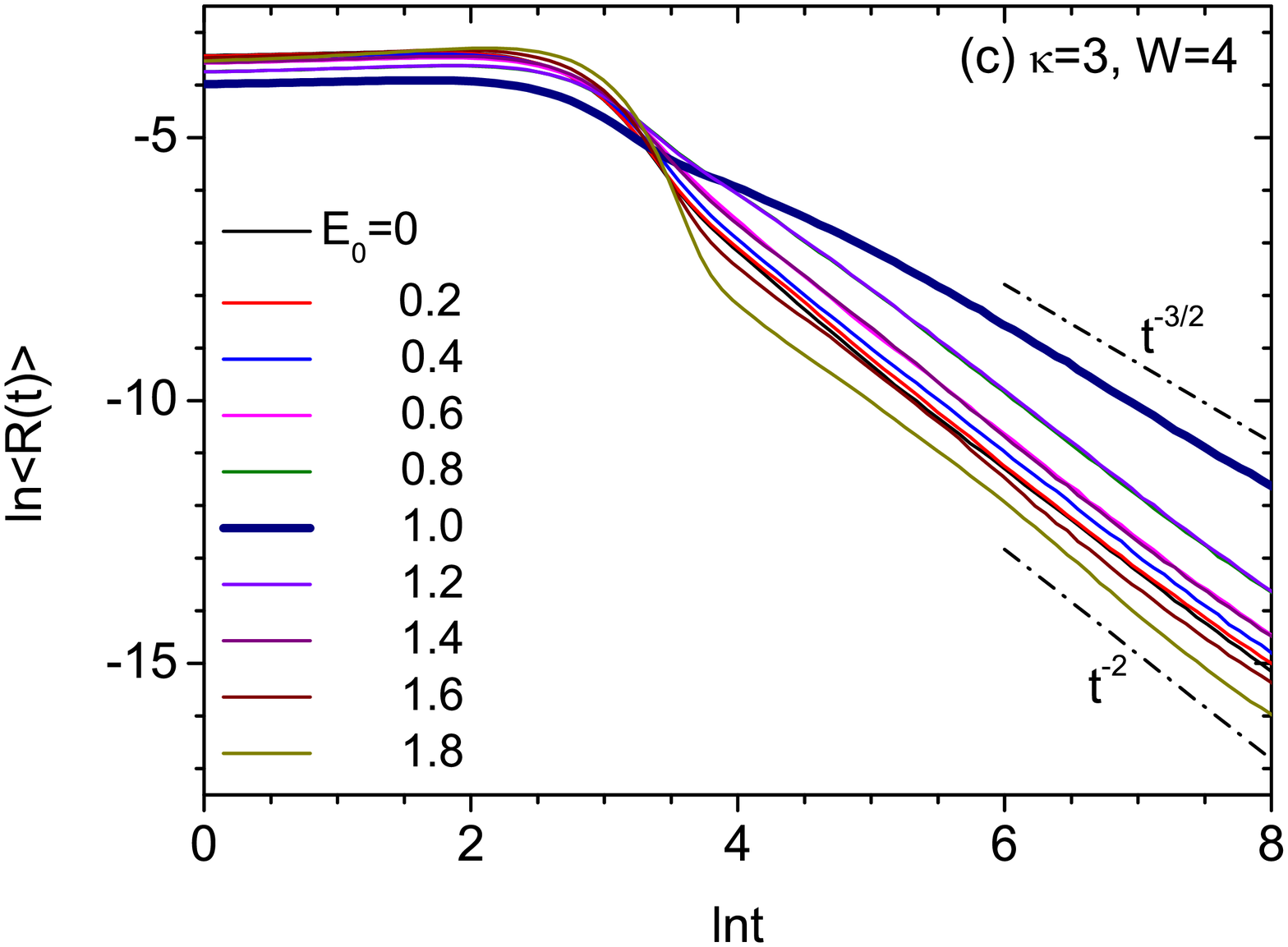}
\caption{Ln--ln plots of $\langle R(t) \rangle$ versus $t$ for various values of $E_{0}$ when $\kappa=3$ and (a) $W=1$, (b) $W=2$, and (c) $W=4$.
$\sigma$, $V_0$, and $L$ are fixed to 0.05, 0, and 1050 respectively. The curve for $E_0=1$ that is highlighted with a thick line scales as $t^{-3/2}$ in the long-time limit in contrast with other curves that scale as $t^{-2}$.}
\label{fig3}
\end{figure}

\begin{figure}
\centering
\includegraphics[width=10cm]{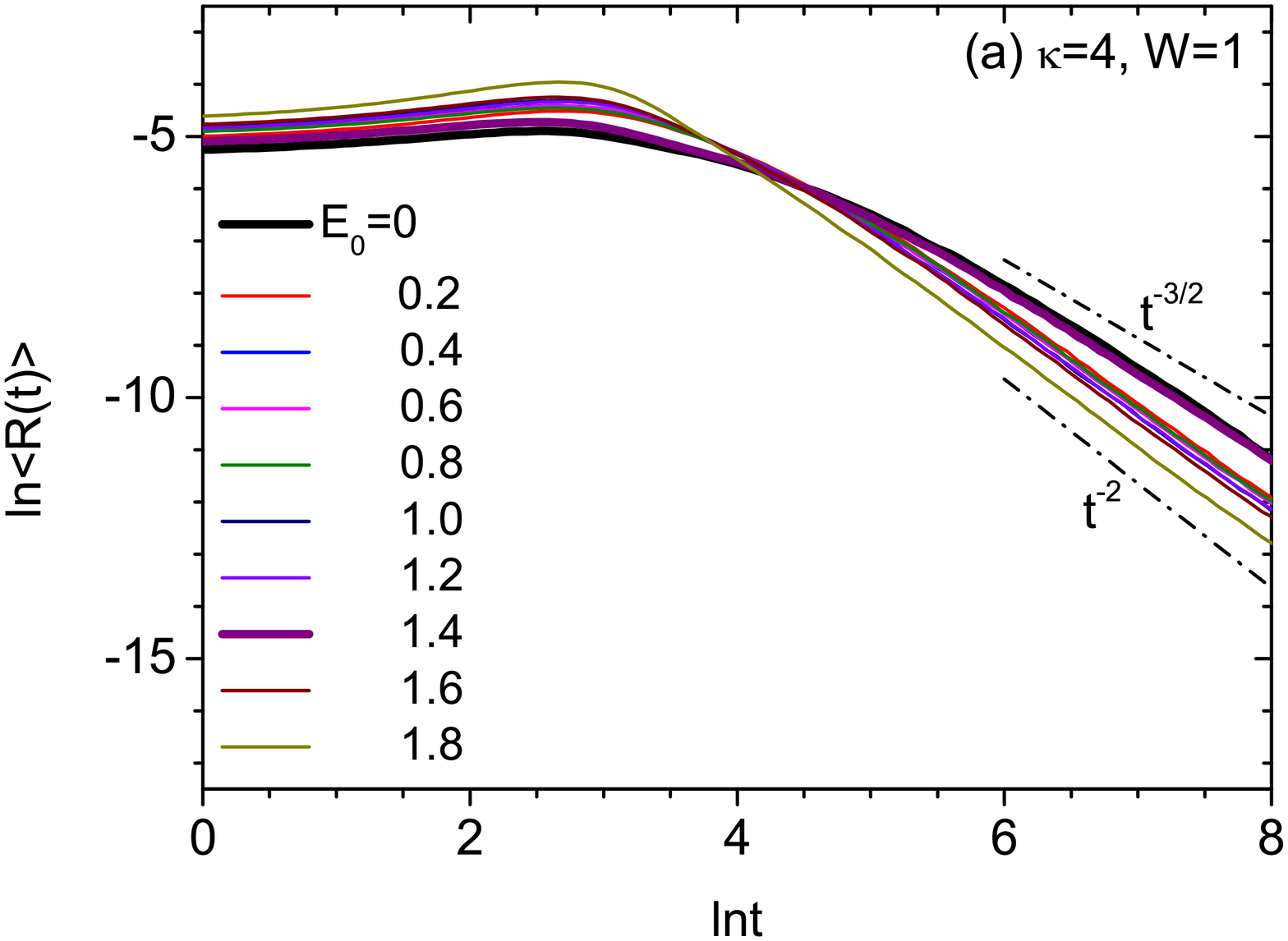}
\includegraphics[width=10cm]{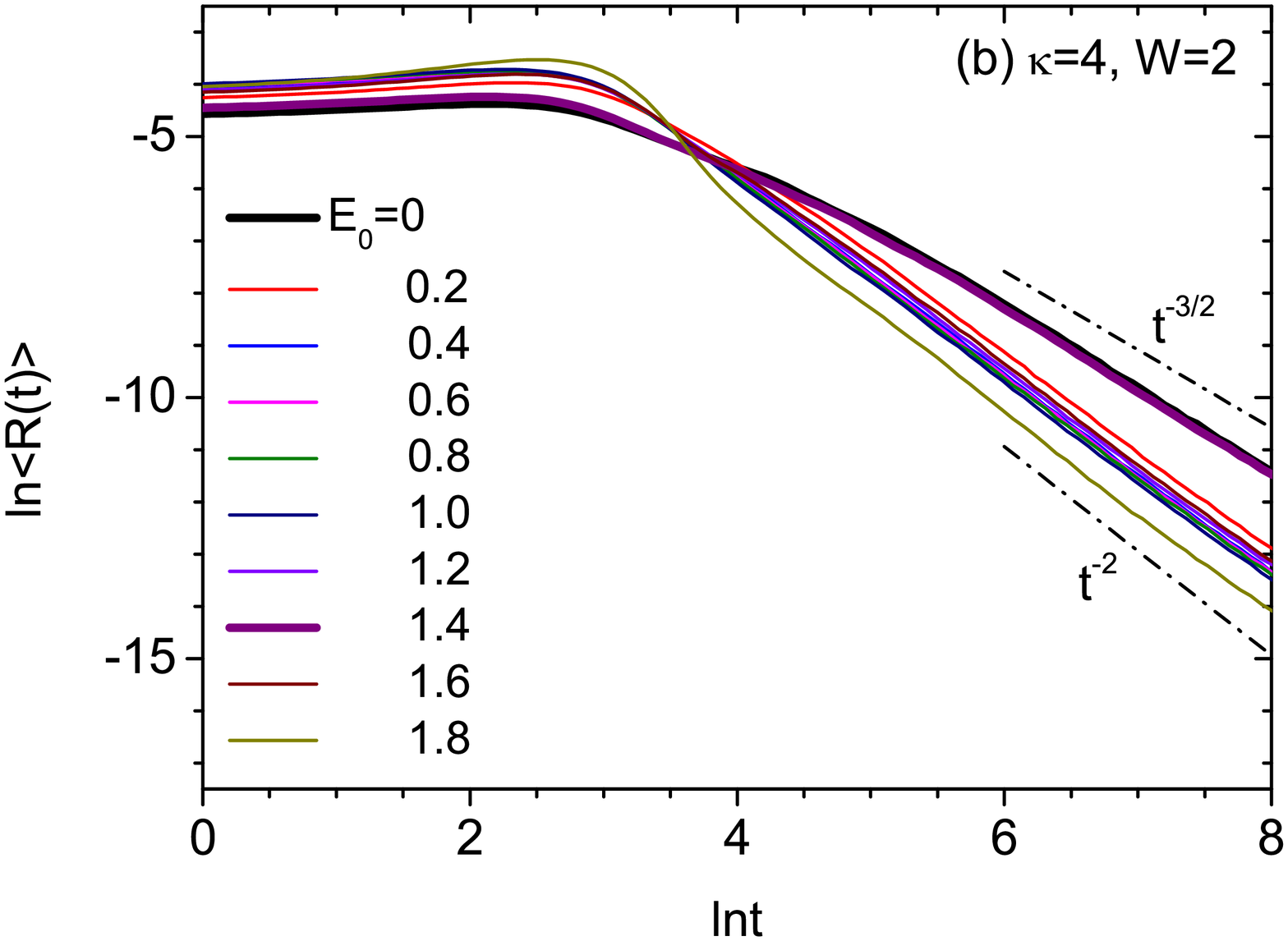}
\includegraphics[width=10cm]{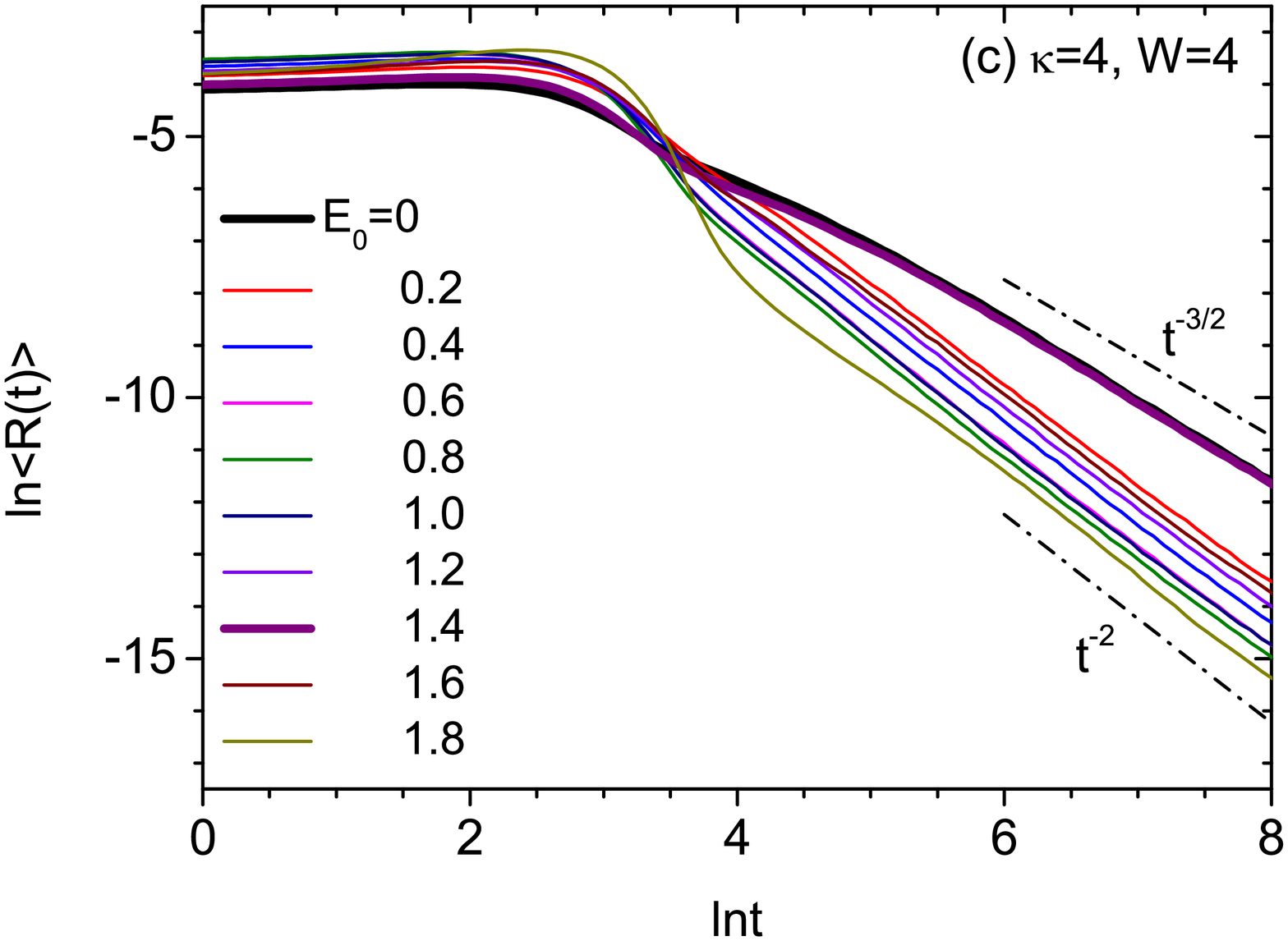}
\caption{Ln--ln plots of $\langle R(t) \rangle$ versus $t$ for various values of $E_{0}$ when $\kappa=4$ and (a) $W=1$, (b) $W=2$, and (c) $W=4$.
$\sigma$, $V_0$, and $L$ are fixed to 0.05, 0,  and 1200 respectively. The curves for $E_0=0$ and 1.4 ($\approx \sqrt{2}$) scale as $t^{-3/2}$ in the long-time limit in contrast with other curves that scale as $t^{-2}$.}
\label{fig4}
\end{figure}
\begin{figure}
\centering
\includegraphics[width=10cm]{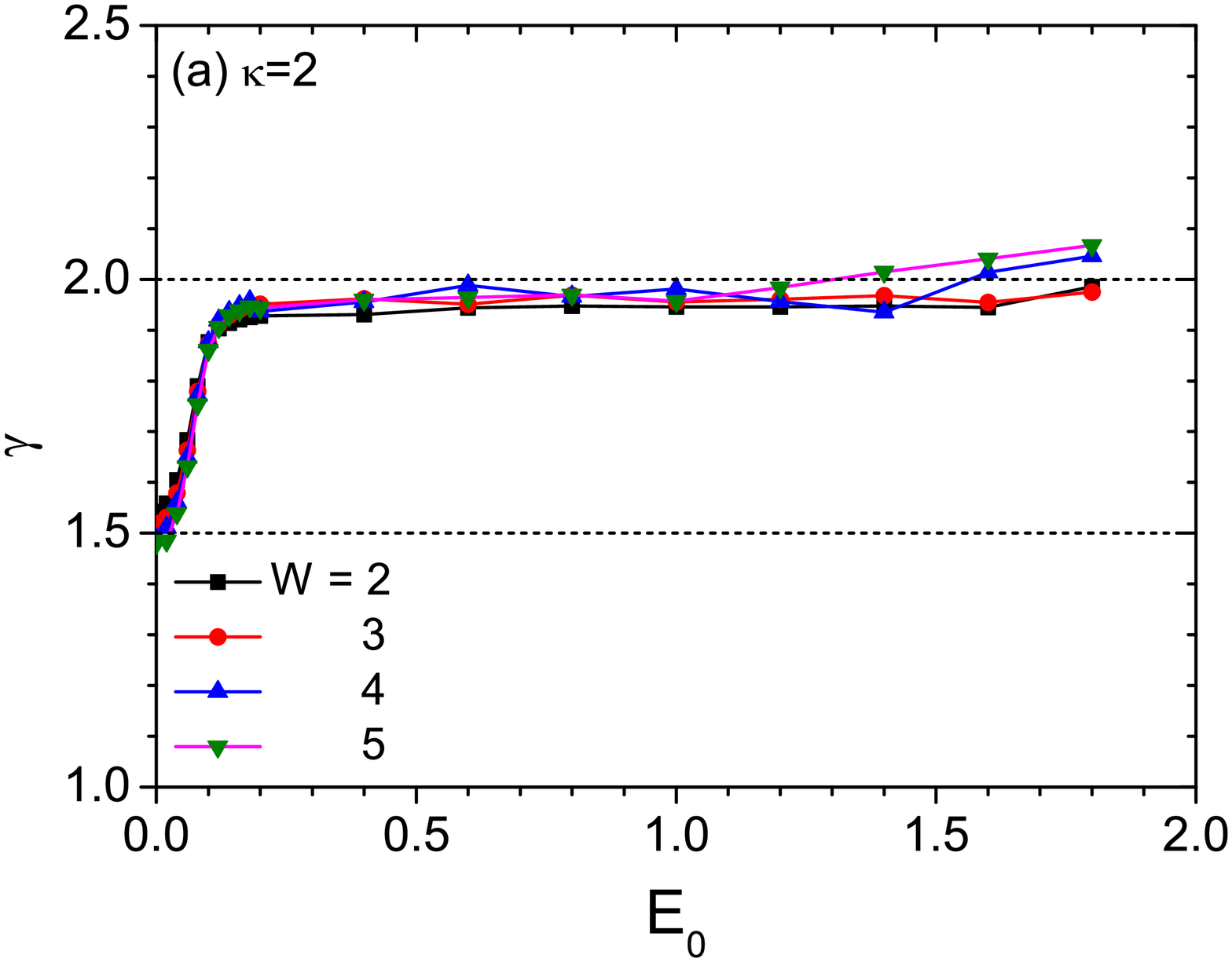}
\includegraphics[width=10cm]{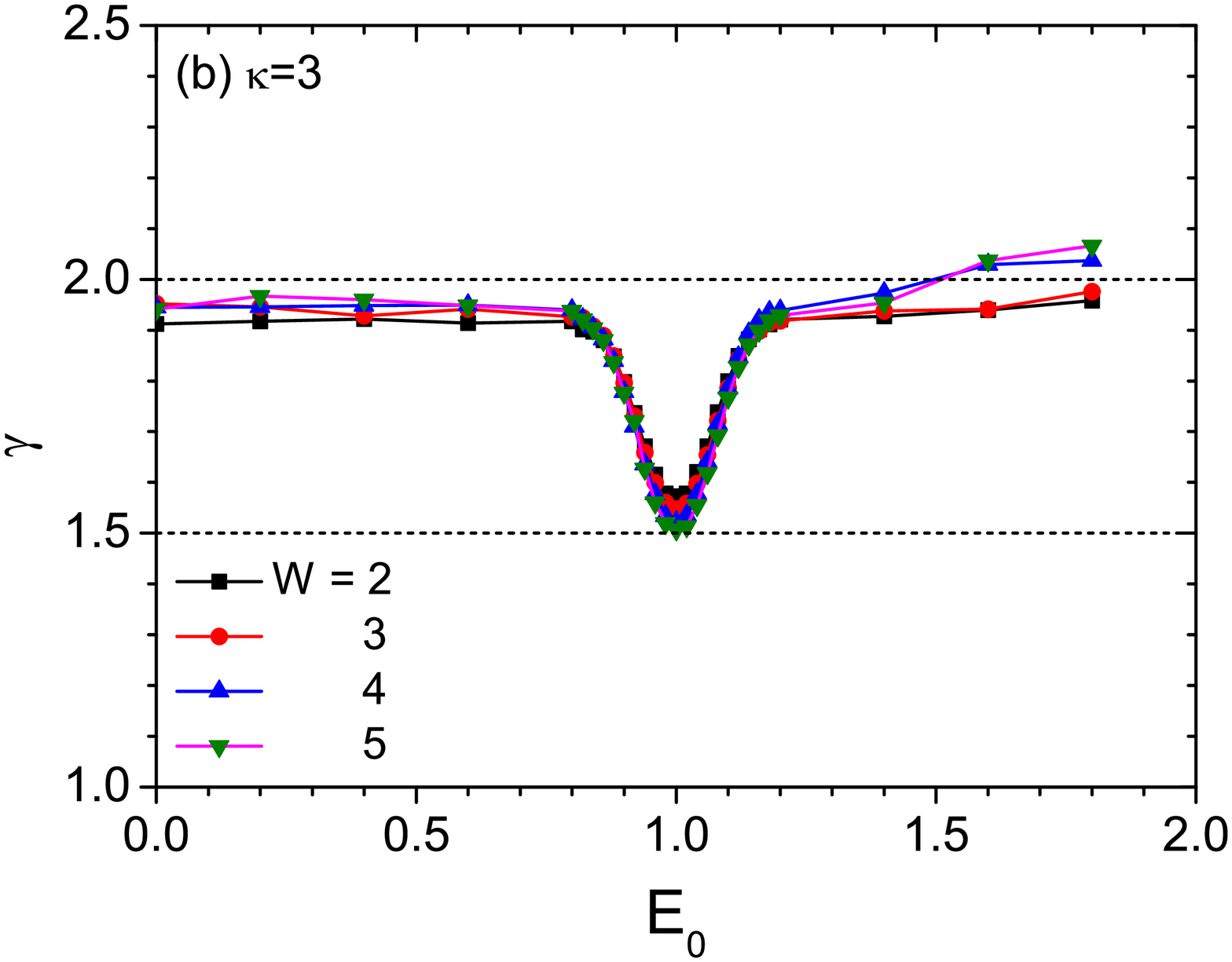}
\includegraphics[width=10cm]{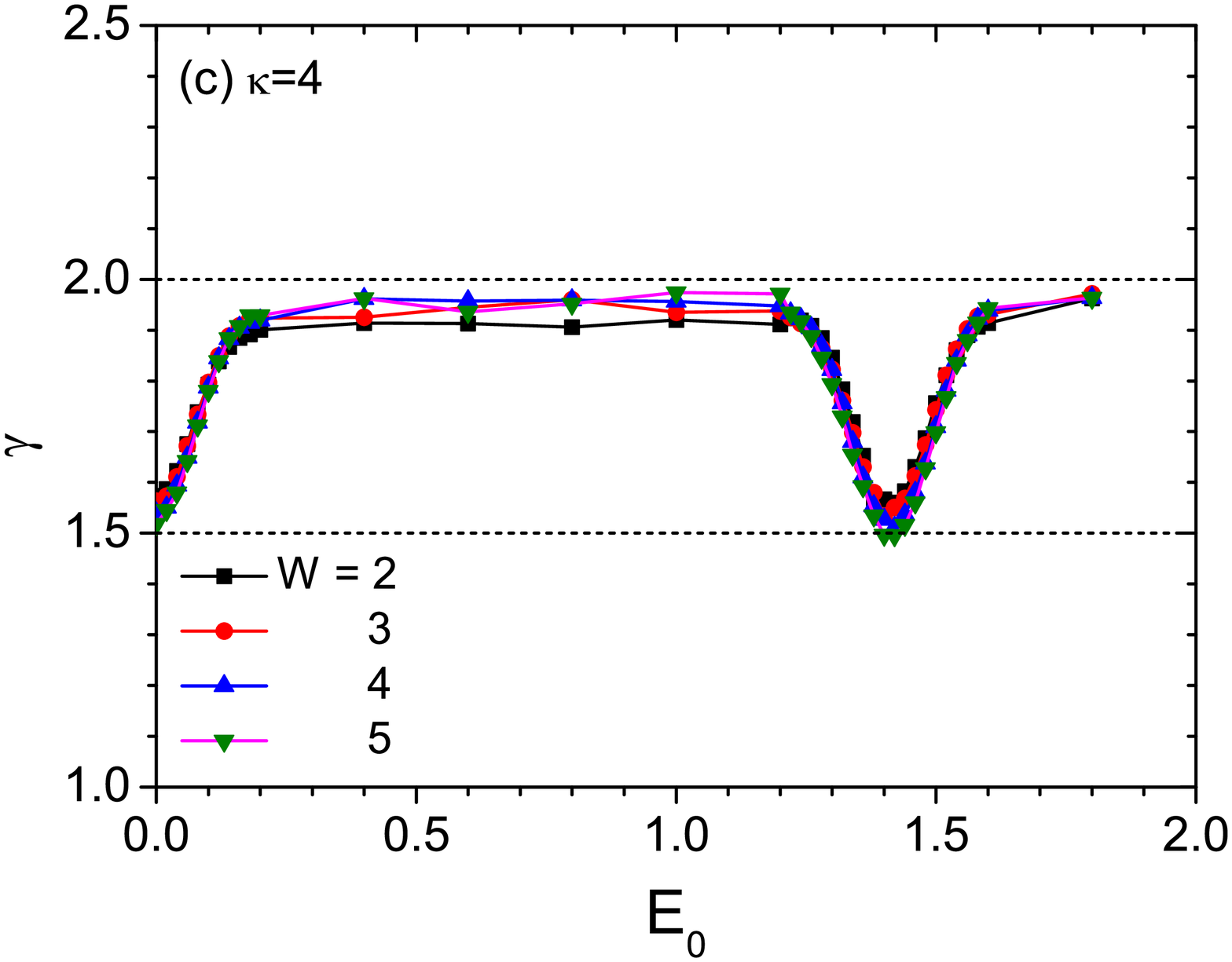}
\caption{The power-law exponent $\gamma$ obtained by fitting the numerical results in the long-time
region where $5<\ln t<8$ plotted versus $E_0$ when $W=2$, 3, 4, 5 and (a) $\kappa=2$, (b) $\kappa=3$, and (c) $\kappa=4$.
$\sigma$ and $V_0$ are fixed to 0.05 and 0 respectively and $L$ is 1000 in (a), 1050 in (b), and 1200 in (c).
The two dashed lines correspond to $\gamma=3/2$ and $2$ which are expected for diffusive and localized wave packets respectively.}
\label{fig5}
\end{figure}

In Fig.~\ref{fig3}, we set $\kappa=3$ and plot the time evolution of $\langle R(t) \rangle$ for the same values of $E_{0}$ and $W$ as in Fig.~\ref{fig2}.
Similarly to the $\kappa=2$ case,
the localization behavior takes place for most values of $E_{0}$, while the diffusive behavior appears only near a special value of $E_{0}$. In contrast to the previous case, however, the diffusive behavior with $\gamma=3/2$ is observed at $E_{0}=1$.
In addition, we have verified numerically that a similar diffusive behavior also occurs at $E_{0}=-1$, though it has not been shown here explicitly.

Next, in Fig.~\ref{fig4}, we consider the $\kappa=4$ case. The overall long-time behavior is similar to the $\kappa=2$ and 3 cases, but the diffusive behavior with $\gamma=3/2$ occurs
at two different values $E_0=0$ and 1.4 ($\approx \sqrt{2}$) in the present case. In addition, a similar behavior is also observed at $E_0=-\sqrt{2}$.
Since the diffusive long-time behavior appears only in a narrow range of $E_0$ values, it can be considered as a kind of quasi-resonance.
Combining the results obtained for $\kappa=2$, 3, and 4, we deduce that the quasi-resonances occur symmetrically with respect to $E_0=0$ and their total number is equal to $\kappa-1$.
We also find that this behavior is unaffected by the disorder strength when $W\gtrsim 1$.

In order to examine the quasi-resonant nature of the diffusive long-time behavior more clearly,
we plot the exponent $\gamma$ obtained by fitting the numerical results in the long-time
region where $5<\ln t<8$ versus $E_0$, when $W=2$, 3, 4, 5 and $\kappa=2$, 3, 4 in Fig.~\ref{fig5}. We point out that there is a mirror symmetry
in a statistical sense with respect to $E_0=0$,
though the region with $E_0<0$ is not shown here. Numerical results for many $E_0$ values around the sharp dips at  $E_{0}=0$, $1$, and $\sqrt{2}$
in addition to those shown in Figs.~\ref{fig2}, \ref{fig3}, and \ref{fig4} have been used in Fig.~\ref{fig5}.
As $E_0$ varies away from the values at the dips, $\gamma$ increases rapidly from 1.5 to 2. The half-widths (full widths at half maximum) of all the dips are similar and roughly equal to 0.15.
In order to show the positions of the quasi-resonances more clearly,
some values of $E_{0}$ versus $\gamma$ around the sharp dips are listed in Table~\ref{table1}.

\begin{table}
\caption{List of the values of $E_{0}$ versus $\gamma$ around the sharp dips at $E_0=1$ in Fig.~\ref{fig5}(b) and at $E_0=1.4$ in Fig.~\ref{fig5}(c). The minimum value of $\gamma$ approaches $3/2$ at $E_{0}=1$ when $\kappa=3$ and at $E_{0}=1.42$ ($\approx\sqrt{2}$) when $\kappa=4$.}
\label{table1}
\begin{ruledtabular}
\begin{tabular}{ c l c c c c }
&~~~~~~~~~~$\kappa=3$&&~~~~~~~~~~~~~~~~~~~~$\kappa=4$\\
\colrule
& $E_{0}$ & $\gamma$ & $E_{0}$ & $\gamma$\\
\colrule
& {0.90} & {1.778} & {1.30} & {1.822}\\
& {0.92} & {1.710} & {1.32} & {1.756}\\
& {0.94} & {1.636} & {1.34} & {1.680}\\
& {0.96} & {1.577} & {1.36} & {1.610}\\
& {0.98} & {1.534} & {1.38} & {1.554}\\
& {\bf 1.00} & {\bf1.521} & {1.40} & {1.527}\\
& {1.02} & {1.535} & {\bf1.42} & {\bf1.510}\\
& {1.04} & {1.577} & {1.44} & {1.536}\\
& {1.06} & {1.640} & {1.46} & {1.580}\\
& {1.08} & {1.715} & {1.48} & {1.638}\\
& {1.10} & {1.786} & {1.50} & {1.710}\\
\end{tabular}
\end{ruledtabular}
\end{table}

\begin{figure}
\includegraphics[width=10cm]{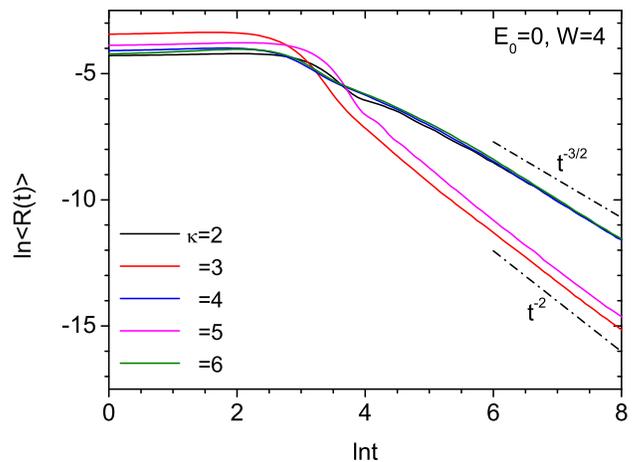}
\caption{Ln--ln plots of $\langle R(t) \rangle$ versus $t$ for several values of $\kappa$ when $E_{0}=0$ and $W=4$.
$\sigma$ and $V_0$ are fixed to 0.05 and 0 respectively.
The curves for even values of $\kappa$ scale as $t^{-3/2}$ while those for odd values of $\kappa$ scale as $t^{-2}$ in the long-time limit.}
\label{fig6}
\end{figure}

In the rest of this subsection, we present some additional results which are useful in deducing a general formula for the quasi-resonant energy values.
In Fig.~\ref{fig6}, we plot $\langle R(t) \rangle$ versus $t$ at $E_{0}=0$ when $W=4$, $\kappa=2$, 3, 4, 5, and 6, and $V_0=0$. We find that the behavior of the incident wave packet at $E_0=0$ depends on the parity of $\kappa$. When $\kappa$ is odd,
the wave packet exhibits  Anderson localization, while, when $\kappa$ is even, it does a diffusive behavior.

\begin{figure}
\includegraphics[width=10cm]{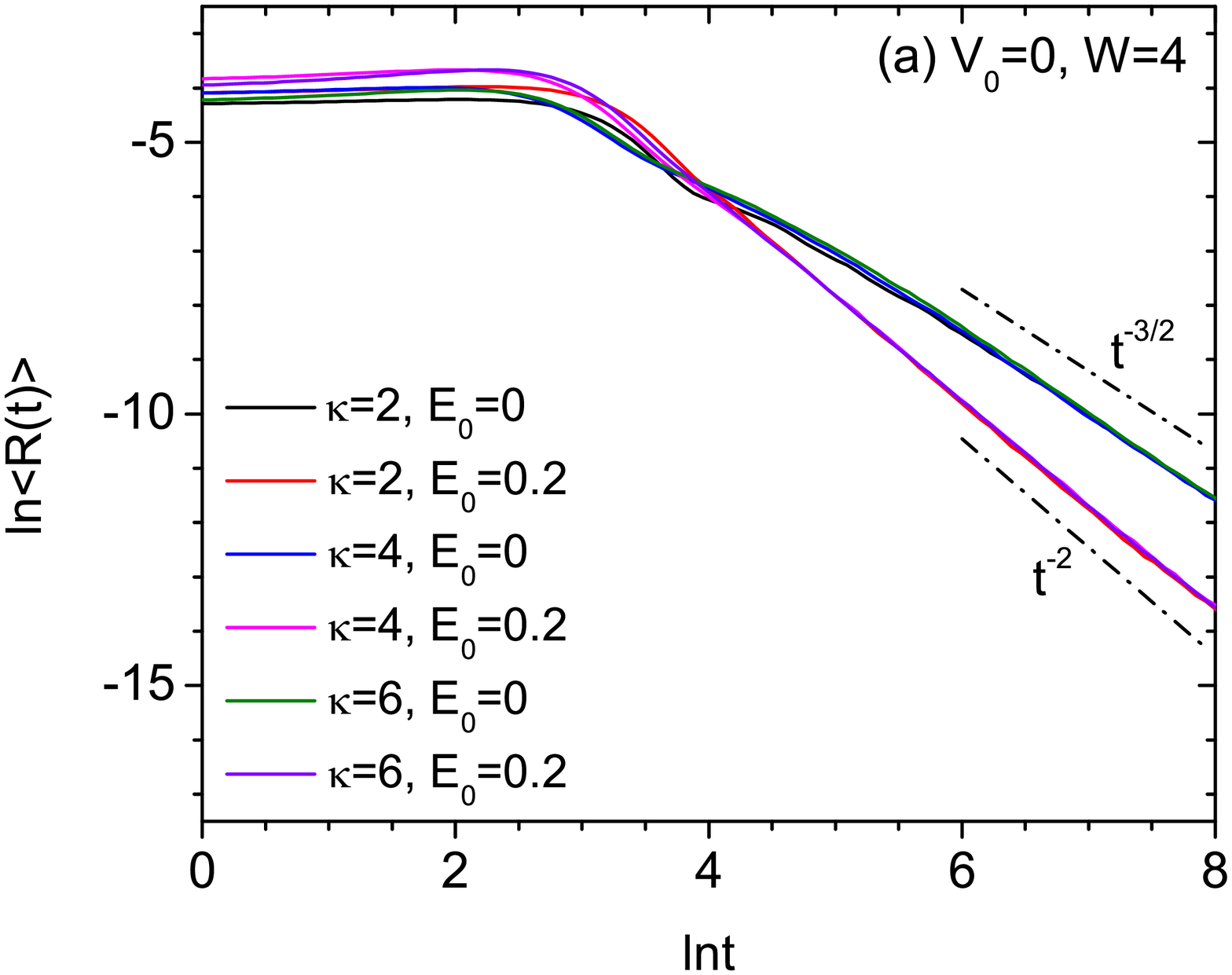}
\includegraphics[width=10cm]{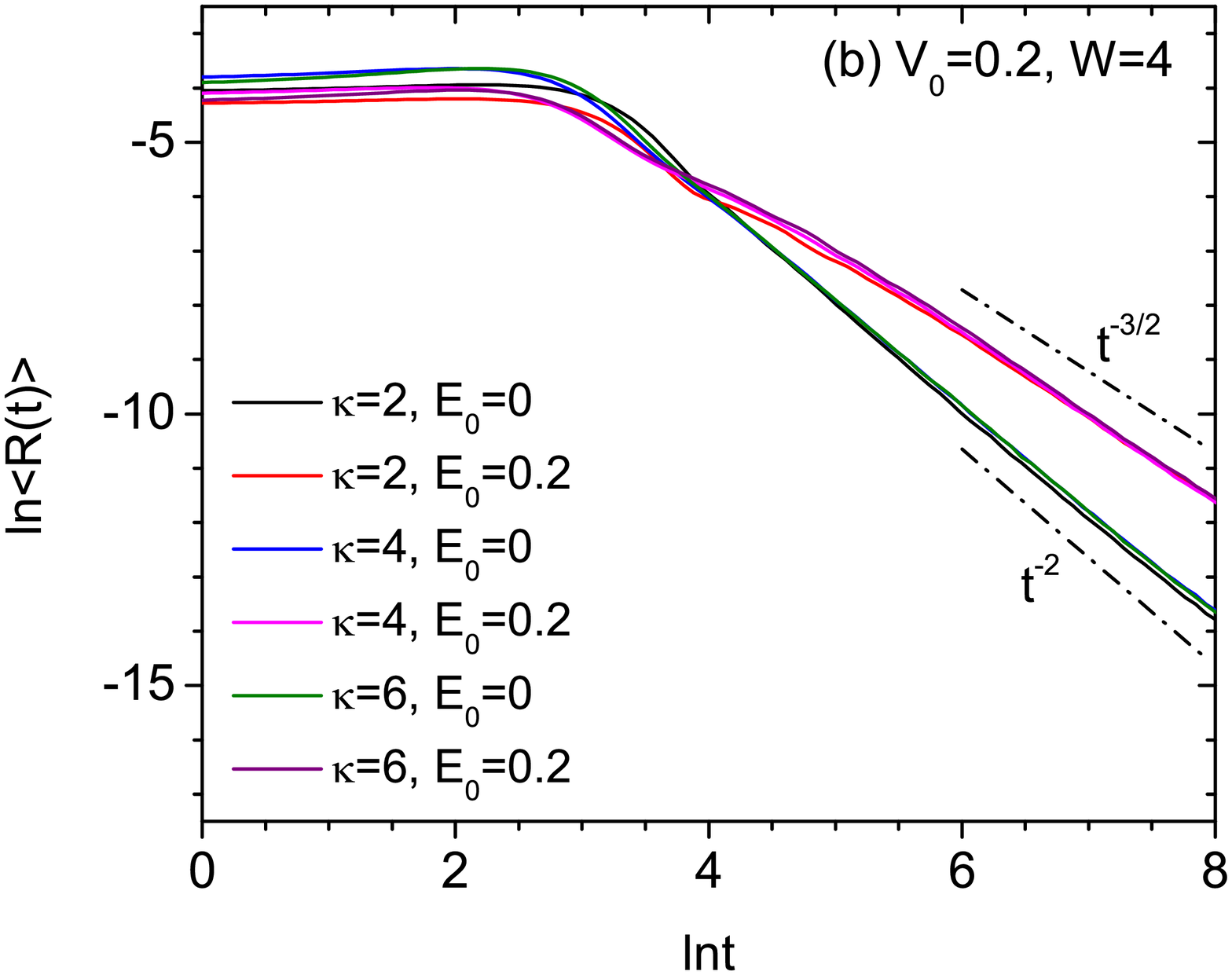}
\caption{Ln--ln plots of $\langle R(t) \rangle$ versus $t$ for $\kappa=2,$ 4, 6, $E_0=0$, 0.2, $\sigma=0.5$, and $W=4$, when (a) $V_{0}=0$ and (b)
$V_0=0.2$. In (a), the curves for $E_0=0$ scale as $t^{-3/2}$ and those for $E_0=0.2$ scale as $t^{-2}$ in the long-time limit.
In (b), the opposite behavior is observed.}
\label{fig7}
\end{figure}

Finally, in Fig.~\ref{fig7}, we compare the long-time scaling behavior at $E_0=0$ and 0.2, when $V_0=0$ and $V_0=0.2$ for $\kappa=2$, 4, and 6.
A diffusive behavior is observed at $E_0=0$ when $V_0=0$, while it is found at $E_0=V_0$ when $V_0=0.2$. Therefore the quasi-resonance energy is found to depend on the value of $V_0$ as well as $\kappa$.

\begin{figure}
\includegraphics[width=10cm]{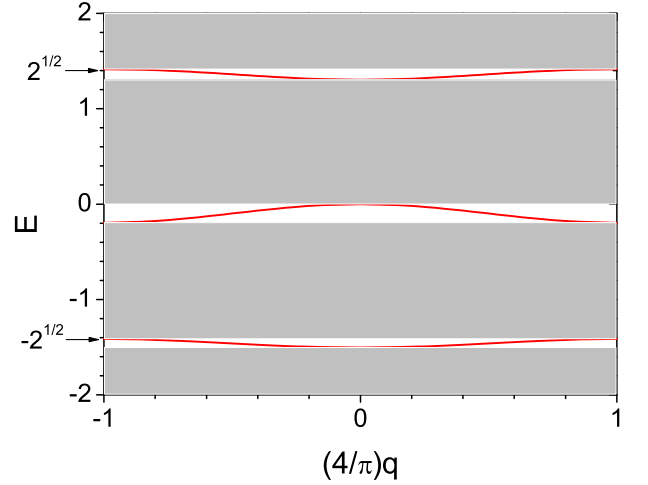}
\caption{Band structure of a periodic mosaic lattice model when $\kappa=4$, $\beta=10$, and $V_0=0$. The gray region denotes the forbidden bands.
The dispersion relation between $E$ and $(4/\pi)q$ is plotted in red inside the allowed bands. The energy values at the upper edges
of the allowed bands are $\sqrt{2}$, 0, and $-\sqrt{2}$, which agree precisely with Eq.~(\ref{eq:res}). }
\label{fig8}
\end{figure}

\subsection{\label{sec:level2} Quasi-resonant diffusion of wave packets}

The numerical results presented above clearly show that a wave packet propagating inside a disordered mosaic lattice displays
a diffusive behavior for some special values of the wave packet's central energy in a quasi-resonant manner, although it exhibits
Anderson localization for all the other values of the central energy. In other words, mosaic modulation of the lattice potential
causes the usual exponential Anderson localization to be destroyed at special discrete values of the energy dependent on the modulation period.
Thus we find that the periodic mosaic modulation is a feature that can induce a new type of delocaliztion in low-dimensional disordered systems.

From our numerical results given in the previous subsection, we can easily deduce that the central energy at the quasi-resonances, $E_R$,
is given precisely by
\begin{eqnarray}
E_{R}=V_{0}+2\cos\left(\frac{\pi}{\kappa}n\right) ~~(n=1,2,\cdots,\kappa-1).
\label{eq:res}
\end{eqnarray}
The number of $E_R$ is equal to $\kappa-1$. $E_R$ is distributed symmetrically with respect to $V_0$ and
$E_R=V_0$ is included only when $\kappa$ is even.
For several small values of $\kappa$, we obtain
\begin{eqnarray}
E_{R}=\left\{\begin{array}{l l}
V_{0} & \quad \mbox{for $\kappa=2$}\\
V_{0}\pm 1 & \quad \mbox{for $\kappa=3$}\\
V_{0}, V_{0}\pm \sqrt{2} & \quad \mbox{for $\kappa=4$}\\
V_0\pm \frac{\sqrt{5}-1}{2}, V_0\pm \frac{\sqrt{5}+1}{2}& \quad \mbox{for $\kappa=5$}\\
V_{0}, V_{0}\pm 1, V_{0}\pm \sqrt{3} & \quad \mbox{for $\kappa=6$}\\ \cdots
\end{array}\right..
\label{equation11}
\end{eqnarray}

The analytical formula for the quasi-resonance energy $E_R$ 
appears to agree with
the formula obtained for the resonance energy [e.g., Eq. (16) of Ref.~29] where the
delocalization occurs in 1D binary random $N$-mer models with $N=\kappa$.
However, this agreement is only coincidental and superficial.
In the binary random $N$-mer model, the resonance energy has been obtained from the condition
that the transmittance through a single $N$-mer is unity, and therefore the transmittance at the resonance energies
is identically equal to 1 and the corresponding 
states are completely extended \cite{Wu,Izr,Maj0,Maj}. In contrast, at our quasi-resonance energies,
the transmittance is not equal to 1 and the states are not extended but critical states, as will be explained in Sec.~\ref{sec:ns}.
In the binary random $N$-mer model of Ref.~29, the on-site potential can take one of the two values 0 and $V_0$.
The problem becomes nonrandom and trivial if $V_0=0$ and so we need to exclude that case.
In order to have extended states, it is also necessary to have the condition that $\vert E_R\vert \le 2$,
which gives another constraint for $V_0$ \cite{Maj0}. On the contrary, in our model, $V_0$ is just a background on-site potential
and can take any arbitrary value including zero.
We also point out that the cosine form for the resonance energy of the $N$-mer model arises only when the on-site potentials at all the $N$ sites of an $N$-mer are the same. 
Even if they are not uniform, there can still exist resonance energies, but they
will not be given by a simple cosine form. 

\begin{figure}
\includegraphics[width=\linewidth]{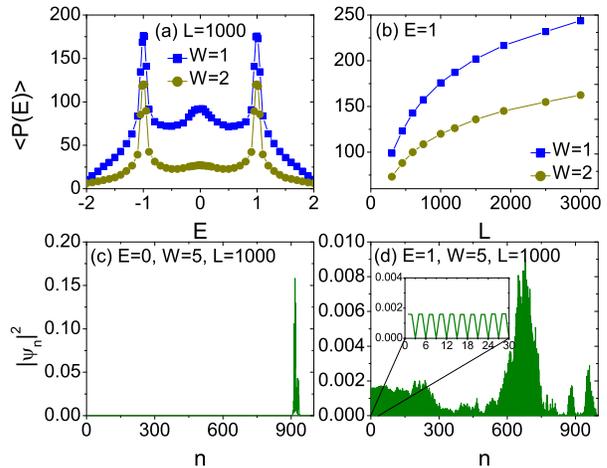}
\caption{Average participation ratio $\langle P(E) \rangle$ of a disordered mosaic lattice  
plotted  versus (a) energy $E$ for $L=1000$ and (b) lattice size $L$ for $E=1$, when $\kappa=3$ and $W=1$, 2. $\langle P(E) \rangle$ is obtained by averaging over the eigenstates within the interval $\Delta E=0.1$ around $E$ and over $1000$ distinct disorder realizations. Two typical wave functions corresponding to the eigenstates with (c) $E=0$ and (d) $E=1$
for a single disorder configuration plotted versus site index $n$ when $\kappa=3$, $L=1000$, and $W=5$.
Inset of (d) shows an expanded plot of the wave function in the range of $0<n<30$.}
\label{fig_supp}
\end{figure}

We pay close attention to the results that the dependence of $\gamma$ on $E_0$ shown in Fig.~\ref{fig5} is insensitive to the strength of
disorder $W$.
In our model, the random potential is present only at periodically spaced sites $n=\kappa m$ ($m=1,2,3,\cdots$). In order for the numerical results
to be insensitive to the disorder strength,
it is natural to assume that the wave-function amplitudes take very small values close to zero at those sites.
That is, {\it the wave function has nodes there}. This implies that the eigenfunctions
have a periodicity
with the wavelength $\lambda$ satisfying
\begin{eqnarray}
\kappa=\frac{\lambda}{2}n~~~~ (n=1,2,3,\cdots).
\end{eqnarray}
For such eigenfunctions, the energy eigenvalues should be very close to the values obtained in the disorder-free case, which are given by
$E=V_0+2\cos q$. Since the wave vector $q$ is related to $\lambda$ by $q=2\pi/\lambda$, we can obtain Eq.~(\ref{eq:res})
in a straightforward way.
This argument strongly suggests that the states at the energies $E_R$ are rather insensitive to the disorder
and therefore are not standard exponentially localized states. Our conjecture that the wave functions
at the quasi-resonance energies have nodes at $n=\kappa m$ will be confirmed later in Fig.~\ref{fig_supp} in
Sec.~\ref{sec:ns}. 

It is highly instructive to consider a related periodic model, which we may call periodic mosaic lattice model, defined by Eq.~(\ref{equation2}), but with
$\beta_n$ replaced by a constant $\beta$. Since this model is strictly periodic with a period $\kappa$, it contains alternating forbidden and allowed bands. In Fig.~\ref{fig8}, we show its band structure when $\kappa=4$, $\beta=10$, and $V_0=0$.
The gray-colored region denotes the forbidden bands.
The dispersion relation between $E$ and $(4/\pi)q$ is plotted in red inside the allowed bands.
We have found that the energy values at the {\it upper edges}
of the allowed bands are precisely given by Eq.~(\ref{eq:res}) for all values of $\kappa$, $\beta$, and $V_0$. As the potential strength $\beta$
increases to large values, the widths of the allowed bands become very small. In the large $\beta$ limit, the allowed bands consist of infinitesimally narrow regions at
the energies given by Eq.~(\ref{eq:res}). If we compare this limiting case with the disordered mosaic lattice model for
large $W$, we find that the forbidden
bands and the narrow allowed bands at $E_R$ in the periodic case are replaced respectively by exponentially localized states and diffusive states in the random case.


\subsection{\label{sec:ns} Nature of the states at the quasi-resonance energies}

In order to better understand the nature of the states at the quasi-resonance energies, we digress briefly from the
study of the time-dependent reflectance to consider the spectral properties of a finite disordered mosaic lattice.
Specifically we calculate the participation ratio $P(E_k)$ for the $k$-th eigenstate with the energy eigenvalue $E_k$,
which is defined by \cite{Thou}
\begin{eqnarray}
P\left(E_{k}\right)=\frac{\left(\sum_{n}\left\vert\psi_{n}^{(k)}\right\vert^2\right)^2}{\sum_{n}\left\vert\psi_{n}^{(k)}\right\vert^4},
\label{eq:pr}
\end{eqnarray}
where $\psi_{n}^{(k)}$ is the value of the $k$-th eigenfunction at the site $n$ ($n=1,2,\cdots,L$).
For a finite lattice, $P(E_k)$ gives approximately the number of sites over which the $k$-th eigenfunction is extended.
To study the large-$L$ scaling behavior of the participation ratio in a disordered system,
it is convenient to introduce a double-averaged quantity $\langle P(E) \rangle$,
where $P(E)$ is obtained by averaging over all eigenstates within a narrow interval $\Delta E$ around $E$
and $\langle\cdots\rangle$ denotes averaging
over a large number of independent disorder configurations.
When $L$ is sufficiently large, we find a power-law scaling behavior of the form
\begin{eqnarray}
\langle P(E) \rangle \propto L^{x}
\label{eq:sca}
\end{eqnarray}
with the scaling exponent $x$.
It has been well-established that for extended states, the scaling exponent $x$ is equal to 1 and
$\langle P(E) \rangle$
increases linearly with increasing the lattice size $L$. On the contrary, for (exponentially) localized states, the exponent $x$
is zero, that is, $\langle P(E) \rangle$ does not depend on $L$ and converges to a constant value as $L\to \infty$.
For critical states at the boundary between extended and localized states,
the exponent should be in the intermediate range of $0<x<1$. Thus the finite-size scaling analysis of $\langle P(E) \rangle$
provides a very useful information about the nature of the states.

In Figs.~\ref{fig_supp}(a) and \ref{fig_supp}(b),
we show the results of numerical calculations of $\langle P(E) \rangle$ obtained when $\kappa=3$, $\Delta E=0.1$,
$W=1$, 2, and the number of disorder configurations is 1000.
In Fig.~\ref{fig_supp}(a), we find that the average participation ratio for $L=1000$
has pronounced peaks at the quasi-resonance energies
$E=\pm 1$, while it remains small at other energies.
We have confirmed that for the value of $E$ away from $\pm 1$, $\langle P(E) \rangle$ approaches a constant
as $L$ increases to large values, implying that $x$ is zero and the states are exponentially localized.
In contrast, at $E=\pm 1$, $x$ is neither 0 nor 1
as demonstrated for $E=1$ in Fig.~\ref{fig_supp}(b). From fitting the data to Eq.~(\ref{eq:sca}), 
we obtain $x\approx 0.25$ for $W=1$ and $x\approx 0.23$ for $W=2$. 
These results clearly demonstrate that the states at the quasi-resonance energies are neither extended nor exponentially localized states. 
We notice that they closely resemble the critical 
states observed at Anderson localisation-delocalisation and quantum Hall plateau transitions \cite{Eve,Cast}.
They are totally different from the resonant states appearing in the binary random $N$-mer model, where such states have been found to be completely extended.

In Figs.~\ref{fig_supp}(c) and \ref{fig_supp}(d), we illustrate the spatial distributions of the wave function amplitudes
obtained for a single disorder configuration when $\kappa=3$, $L=1000$, $W=5$, and $E=0$, 1.
When the energy is away from the quasi-resonance energies, we find that the wave function is strongly localized
as in Fig.~\ref{fig_supp}(c). At the quasi-resonance energies, however, the wave function
has a distinct spatial distribution with several disjointed occupied regions as in Fig.~\ref{fig_supp}(d).
This is a unique characteristic frequently observed in critical or multifractal states.
We also find that the wave function amplitudes are very close to zero
at all the sites satisfying $n=m\kappa$ ($m=1,2,\cdots$) as shown in the inset of Fig.~\ref{fig_supp}(d), which is fully consistent with our conjecture that there are
wave function nodes at such sites. This node structure is absent for localized wave functions
and occurs only at the quasi-resonance energies.
It leads to the behavior that when $W$ is sufficiently large, the quasi-resonant states
are insensitive to the strength of disorder.

From separate calculations for the eigenfunctions of the periodic mosaic model,
we have confirmed that they also have nodes at $n=m\kappa$ ($m=1,2,\cdots$) only for the energies given by Eq.~(\ref{eq:res}).
Let us suppose that starting from the periodic mosaic model, 
we turn on the weak disorder at the sites $n=m\kappa$ ($m=1,2,\cdots$) 
and increase its strength gradually. The special node structure of the wave function we have discussed so far 
will be maintained at the quasi-resonance energies regardless of the strength of disorder. 
Therefore the node structure of the wave functions is a crucial feature that connects the periodic 
and disordered mosaic lattice models and induces the critical states at the quasi-resonance energies.
A more systematic and comprehensive analysis about the nature
of the quasi-resonant states in the framework of the spectral problem will be presented elsewhere. 


\section{\label{sec:level1} Conclusion}

In this paper, we have studied numerically the time evolution of Gaussian wave packets in an effectively semi-infinite disordered mosaic lattice chain where the on-site potential takes a random value only at periodically spaced sites. We have performed extensive numerical calculations of the disorder-averaged time-dependent reflectance, $\langle R(t) \rangle$, for various values of the wave packet's central energy $E_0$,
the modulation period $\kappa$, and the disorder strength $W$. We have found that
the long-time behavior of $\langle R(t)\rangle$ obeys a power-law decay of the form $t^{-\gamma}$ in all cases.
In the absence of the mosaic modulation (i.e., $\kappa=1$), the exponent $\gamma$ is equal to 2 regardless of the parameters, implying
the onset of the standard Anderson localization. When the mosaic modulation is turned on (i.e., $\kappa\ge 2$),
$\gamma$ is still equal to 2 for almost all values of $E_0$, while at a finite number (equal to $\kappa-1$) of discrete values
of $E_0$ dependent on $\kappa$, $\gamma$ approaches 3/2, implying the onset of the classical diffusion.
We have found that this phenomenon is independent of $W$ as long as it is sufficiently large and occurs in a quasi-resonant manner such that $\gamma$ varies rapidly from 3/2 to 2
in a narrow energy range as $E_0$ varies away from the quasi-resonance values.
We have deduced a simple analytical formula for the quasi-resonance energies and provided an explanation of this novel delocalization phenomenon based on the interplay between randomness and band structure
and the node structure of the wave functions.
We have also explored the nature of the states at the quasi-resonance energies using a finite-size scaling analysis
of the average participation ratio and found that the states
are neither extended nor exponentially localized, but ciritical states. 
The model proposed here can be readily realized experimentally using various physical systems, which include coupled optical waveguide arrays,
synthetic photonic lattices, and ultracold atoms.
 
In the future research, it is desirable to perform a comprehensive spectral analysis of the participation ratio, the logarithmic transmittance,
and the localization length to better elucidate the true nature of the states
at the quasi-resonance energies as well as to test the theoretical predictions experimentally. 
Another interesting direction of research is to explore the spreading dynamics of a wave packet 
launched at the center of a long disordered mosaic lattice, which is expected to provide an understanding of 
the states at the quasi-resonance energies from a different perspective. 

\begin{acknowledgments}
B.P.N. would like to thank Felix Izrailev for carefully reading a draft version of the manuscript and providing valuable comments.
We also appreciate greatly very helpful comments and suggestions by an anonymous referee and Seulong Kim.  
This research was supported through a National Research
Foundation of Korea Grant (NRF-2022R1F1A1074463)
funded by the Korean Government.
It was also supported by the Basic Science Research Program funded by the Ministry of Education (2021R1A6A1A10044950) and by
the Global Frontier Program (2014M3A6B3063708).
\end{acknowledgments}

\newpage

\end{document}